\documentclass[aps,prl,twocolumn,showpacs,groupedaddress]{revtex4-1}
\usepackage{amsmath}
\usepackage{amssymb}
\usepackage{graphicx}
\usepackage{dcolumn}
\usepackage{bm}

\begin{document}

\title{Stability of resonantly interacting heavy-light Fermi mixtures}
\author{G. E. Astrakharchik$^{\rm a}$, S. Giorgini$^{\rm b}$, and J. Boronat$^{\rm a}$}
\affiliation{
$^{\rm a}$ Departament de F\'{i}sica i Enginyeria Nuclear, Universitat Polit\`{e}cnica de Catalunya, Campus Nord B4-B5, E-08034, Barcelona, Spain \\
$^{\rm b}$  Dipartimento di Fisica, Universit\`a di Trento and INO-CNR BEC Center, I-38050 Povo, Trento, Italy}

\begin{abstract}
We investigate a two-component mixture of resonantly interacting Fermi gases as a function of the ratio $\kappa$ of the heavy to the light mass of the two species. The diffusion Monte Carlo method is used to calculate the ground-state energy and the pair correlation function starting from two different guiding wave functions, which describe respectively the superfluid and the normal state of the gas. Results show that the mixture is stable and superfluid for mass ratios smaller than the critical value $\kappa_c = 13\pm1$. For larger values of $\kappa$ simulations utilizing the wave function of the normal state are unstable towards cluster formation. The relevant cluster states driving the instability appear to be formed by one light particle and two or more heavy particles within distances on the order of the range of the interatomic potential. The small overlap between the wave function of the trimer bound state and the guiding wave function used to describe the superfluid state produces the unphysical stability of the superfluid gas above $\kappa_c$.
\end{abstract}

\maketitle

Two-component mixtures of fermionic gases with resonant interspecies interactions constitute a fundamental paradigm in the physics of ultracold gases. In the last decade a large amount of experimental and theoretical work has been devoted to the study of mixtures of two different hyperfine states in a single species gas (either $^6$Li or $^{40}$K) in the close vicinity of a Fano-Feshbach resonance~\cite{review}. At low enough temperature and for balanced populations of the two components, the gas has been found to be superfluid irrespective of the strength of interactions. The pairing mechanism, however, changes as a function of coupling from the Cooper pairing in the Bardeen-Cooper-Schrieffer (BCS) region to the molecule formation in the Bose-Einstein condensate (BEC) regime. The unitary limit, corresponding to the resonance position, has attracted the greatest attention because of its universal character, as the inverse Fermi momentum $k_F^{-1}$ and the Fermi energy $\epsilon_F$ are the only relevant length and energy scale, respectively.

More recent important experimental advances concern the realization of resonantly interacting mixtures with two different fermionic species~\cite{innsbruck1, munich1, innsbruck2, amsterdam1, munich2, innsbruck3, innsbruck4}. In this case a new degree of freedom, namely the ratio $\kappa=M/m$ of the heavy to the light mass of the two species, enters the problem leading to interesting theoretical predictions for the superfluid ground state~\cite{iskin1, baranov, iskin2}.

The effects of mass asymmetry between the heavy (\textit{h}) and light (\textit{l}) components of the mixture have been thoroughly investigated at unitarity for systems of few atoms, both in free space and in harmonic traps~\cite{efimov, petrov, nishida, castin, blume1, blume2, carlson1}. It has been found~\cite{efimov, petrov} that an infinite number of Efimov states composed by two \textit{h} and one \textit{l} particles appear when the mass ratio $\kappa> 13.6$. In the range $8.62 < \kappa < 13.6$ interactions can be fine tuned to produce triplet bound states~\cite{nishida}. Furthermore, few-body calculations of systems composed by four and five particles indicate that a bound state can appear at $\kappa \simeq 10.4$ for the system (3\textit{h}+1\textit{l}) and at $\kappa \simeq 9.8$  for (4\textit{h}+ 1\textit{l})~\cite{blume2,carlson1}.

Many-body heavy-light Fermi mixtures at unitarity have also been studied using quantum Monte Carlo (QMC) methods by von Stecher \textit{et al.}~\cite{blume3} in harmonic traps and by Gezerlis \textit{et al.}~\cite{carlson2} in homogeneous systems. The aim of Ref.~\cite{carlson2} was mainly to analyze the particular case of $^{40}$K-$^6$Li mixtures with $\kappa \simeq 6.5$ as a function of the population imbalance. Interestingly, they comment that the normal state of the gas collapses to a many-body bound state (cluster) for mass ratios $\kappa \sim 12$. This finding points to the existence at unitarity of a critical value $\kappa_c$ above which the homogeneous gas is mechanically unstable against density fluctuations. The strongly interacting $^{40}$K-$^6$Li mixture, already produced in experiments~\cite{innsbruck1, munich1, innsbruck2, amsterdam1, munich2, innsbruck3, innsbruck4}, has a mass ratio too small for observing this instability and one can think that other possibilities, as for instance $^{173}$Yb-$^6$Li~\cite{hara}, could be brought to the resonant regime in a near future. Moreover, even for lighter species one could also imagine that the mass ratio can be tuned in a nearly continuous way through proper confinement in an optical lattice (in this case, one relies on effective masses)~\cite{baranov}.

In the present work we investigate the stability of heavy-light Fermi mixtures at unitarity by carrying out a fully microscopic study using the diffusion Monte Carlo (DMC) method within the fixed-node (FN) approximation. This technique has become nowadays a standard tool used extensively in many problems in condensed-matter physics including ultracold gases~\cite{mitas}. We use an approach similar to Ref.~\cite{astra} to calculate the ground-state energy and the pair correlation function starting from a guiding wave function that describes either the superfluid or the normal state of the gas. We find that the latter shows an instability towards cluster formation at the critical mass ratio $\kappa_c = 13\pm1$. In the case of the superfluid state, the apparent lack of instability can be understood as an artifact in the nodal surface of the corresponding guiding wave function.

The Hamiltonian of a mixture of heavy and light fermions with masses $M$ and $m$ is written as
\begin{equation}
H=
-\frac{\hbar^2}{2 M} \sum_{i=1}^{N_h} \bm{\Delta}^{\textit{h}}_i
-\frac{\hbar^2}{2 m} \sum_{i=1}^{N_l} \bm{\Delta}^{\textit{l}}_i
+\sum_{i,j=1}^{N_h,N_l} V(|{\bf r}_i^{\textit{h}}-{\bf r}_j^{\textit{l}}|) \ ,
\label{hamiltonian}
\end{equation}
where $\bm{\Delta_i^{h(l)}}$ denotes the Laplacian operator for the heavy (light) particles. Periodic boundary conditions in a cubic box of volume $V=L^3$ are used and we restrict our study to balanced populations, $N_h=N_l=N/2$, $N$ being the total number of particles. Intraspecies interactions are neglected while short-range interspecies interactions are modeled by a square-well potential with depth $V_0$: $V(r)=-V_0$ if $r<R$ and $V(r)=0$ elsewhere. For this potential the $s$-wave scattering length is known analytically: $a= R \left[1-\tan(K_0 R)/(K_0 R)\right]$, with $K_0^2= \mu V_0/\hbar^2$ and $\mu=(M^{-1}+m^{-1})^{-1}$ the reduced mass. Our interest is in the unitary limit where $|a|=\infty$ and the resonance condition gives in our case $K_0 R=\pi/2$, corresponding to the appearance of the first two-body bound state in the well. To ensure universality the value of $R$ is chosen much smaller than the mean interparticle distance determined by the density $n=N/V$: as in Ref.~\cite{astra} we use $nR^3=10^{-6}$.

We carry out simulations using two different guiding wave functions. The first is a Jastrow-Slater (JS) function given by
\begin{equation}
\Psi_T({\bf r}_1,...,{\bf r}_N)
= \left[ \prod_{i,j=1}^{N_h,N_l} f(|{\bf r}_i^\textit{h}-{\bf
r}_j^\textit{l}|) \right] \;\;
\text{det}[\varphi({\bf r}_i^\textit{h}-{\bf r}_j^\textit{l})] \;,
\label{JS}
\end{equation}
where $f(r)$ is a non-negative function of the interparticle distance reproducing at short separations the zero-energy solution of the two-body problem with the potential $V(r)$ and satisfying the boundary condition $f^\prime(r=L/2)=0$ for its derivative at half the box width. The pair orbital in the determinant is given by the combination of plane waves $\varphi({\bf r})=\sum_{n^2\le4}\cos({\bf k}_{{\bf n}}\cdot{\bf r})$, where ${\bf k}_{{\bf n}}=2\pi(n_x,n_y,n_z)/L$ are the wave vectors determined by the integers $n_{x,y,z}=0,\pm1,...$ and the sum is restricted over the filled shells $n^2=n_x^2+n_y^2+n_z^2\le 4$, corresponding to $N=66$ particles used in the simulations. The determinant in (\ref{JS}) fixes the nodal surface of the many-body wave function to that of a non-interacting gas, being therefore incompatible with off-diagonal long-range order. As a consequence, the Jastrow-Slater wave function (\ref{JS}) describes the normal state of the gas. Simulations of the superfluid state are instead performed using the superfluid wave function
\begin{equation}
\Psi_T({\bf r}_1,...,{\bf r}_N)
=\text{det} [\phi(|{\bf r}_i^\textit{h}-{\bf r}_j^\textit{l}|)]\;,
\label{Pair}
\end{equation}
where the pair orbital $\phi(r)$ is chosen  as  the two-body solution, similarly to the Jastrow factor $f(r)$ in Eq.~(\ref{JS}). In fact, Eq.~(\ref{Pair}) corresponds to a BCS wave function characterized by a spherically symmetric order parameter. In the FN-DMC algorithm an important systematic bias concerns the number of walkers $N_w$, \textit{i.e.} points ${\bf R}=({\bf r}_1,...,{\bf r}_N)$ in configuration space where the wave function is sampled. The FN-DMC energies should be estimated in the limit $N_w \rightarrow \infty$. However, if the guiding  function $\Psi_T$ is reasonably accurate, the dependence on $N_w$ can be neglected for some large enough number of walkers. On the contrary, a strong dependence on $N_w$ points out the effect of some relevant physics not accounted for by $\Psi_T$.

Mean-field theory has been applied to the problem of the BCS-BEC crossover at $T=0$ with mass asymmetry~\cite{iskin1}, but the results obtained show that both the chemical potential and the gap parameter, once the Fermi energy is rescaled with the reduced mass $\epsilon_F=\hbar^2k_F^2/4\mu$ where $k_F=(3\pi^2n)^{2/3}$, are unchanged compared to the corresponding quantities of the equal mass problem. Non-trivial effects arising from the mass ratio appear in the expression of the gap parameter if one includes perturbative corrections following the approach by Gorkov and Melik-Barkhudarov~\cite{baranov}.

\begin{figure}
\centerline{
\includegraphics[width=0.8\linewidth]{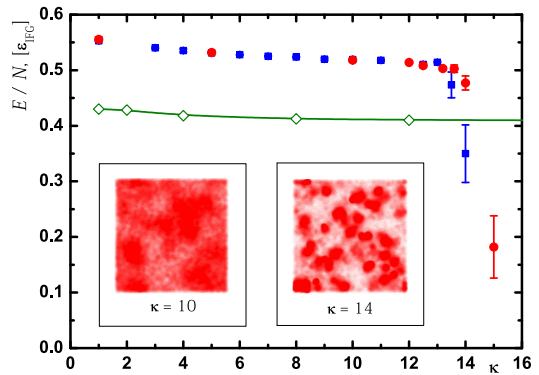}}
\caption{(color online). Energy per particle in units of $\epsilon_{IFG}$ as a function of the mass ratio $\kappa$. Diamonds correspond to the results for the superfluid state. Circles and squares correspond instead to simulations of the normal state with a number of walkers $N_w=1000$ and $5000$, respectively. The insets show for this latter case a two-dimensional projected density for two different values of $\kappa$.}
\label{fig1:kappa}
\end{figure}

We calculate the energy as a function of the mass ratio up to large values of $\kappa$, in particular larger than $\kappa=13.6$ where bound trimers are expected to appear. The FN-DMC energies of the superfluid and normal state are shown in Fig.~\ref{fig1:kappa}. For $\kappa=1$ we find that the superfluid state is lower in energy than the normal state: $\xi=0.42 \pm 0.01$ vs. $\xi=0.54 \pm 0.01$~\cite{astra, carlson3, carlson4, gandolfi}. Here $\xi$ is the proportionality coefficient between the energy per particle of the interacting system $E/N=\xi \epsilon_{IFG}$ and the one of a free Fermi gas $\epsilon_{IFG}=3\epsilon_F/5$. For $\kappa>1$ both energies decrease slightly with increasing $\kappa$, in agreement with the findings of Ref.~\cite{carlson2} reporting for $\kappa=6.5$ a $\sim 5$\% decrease of the energy of the superfluid state compared to the equal mass case. We notice that in the range $\kappa\lesssim 13$ the energy of the normal state lies above the one of the superfluid state implying that the latter is the stable ground-state of the system. For $\kappa>14$ the behavior in the two cases is completely different: the energy calculated using the superfluid guiding wave function slightly decreases with $\kappa$, whereas the energy obtained using the normal gas wave function shows a fast-growing instability towards negative energies. According to the variational nature of fixed-node calculations, we conclude that the gas becomes unstable against cluster formation around $\kappa\simeq13$. This instability is explicitly observed in the particle configurations of the FN-DMC simulation. In the insets of Fig.~\ref{fig1:kappa}, we show typical snapshots of the particle positions for two mass ratios, below and above the instability. For $\kappa=10$ the gas is structureless and homogeneous, whereas for $\kappa=14$ the system has formed clusters indicating a spinodal decomposition.

\begin{figure}
\centerline{
\includegraphics[width=0.8\linewidth,angle=0]{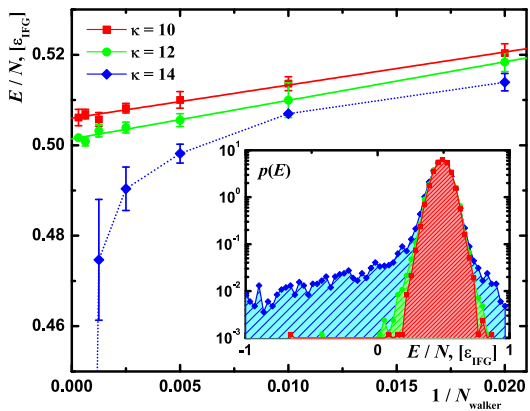}}
\caption{(color online): Energy per particle of the normal state with $N=14$ as a function of the inverse of the number of walkers $N_w$ for $\kappa=10$, $12$, and $14$ (respectively red, green and blue symbols). The inset shows the energy probability distribution for the same values of $\kappa$ and $N_w=800$.}
\label{fig2:walkers}
\end{figure}

The dependence of the energy of the normal state on the number of walkers is first shown as a function of the mass ratio in Fig.~\ref{fig1:kappa} for $N_w=1000$ and $5000$ and is further analyzed in Fig.~\ref{fig2:walkers}. For $\kappa=10$ and $12$ the energy shows a linear decrease with increasing $N_w$ towards the limit $1/N_w = 0$ with a well defined asymptotic value. On the contrary, for $\kappa=14$ the energy decreases fast with increasing $N_w$ and for large $N_w$ we observe negative energies corresponding to the formation of self-bound many-body clusters. The appearance of clusters is also reflected in the large increase of the statistical error when $\kappa=14$ and $N_w$ is large. Additional insight on the increase in variance can be inferred from the probability distribution of local energies. The energy histograms are shown in the inset of Fig.~\ref{fig2:walkers}. For $\kappa=10$ and $12$ there is a well defined Gaussian peak, centered on the mean value, whose width determines the variance of the energy estimate. For $\kappa=14$ the picture is appreciably different: large tails in the distribution are developed, mainly towards lower energies.

\begin{figure}[t]
\centerline{
\includegraphics[width=0.6\linewidth,angle=-90]{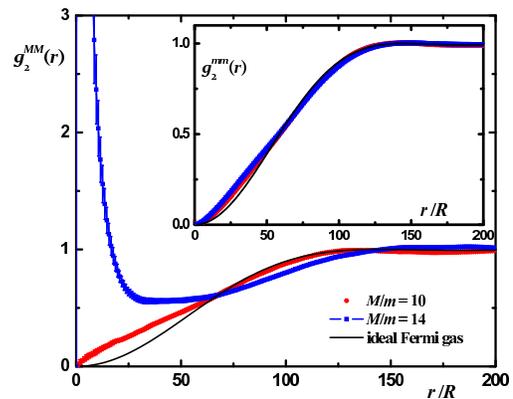}}
\caption{(color online): Pure estimators for the two-body radial distribution functions $g_2^{\text{hh}}(r)$ and $g_2^{\text{ll}}(r)$ for two different values of $\kappa$. Distances are in units of the interaction range $R$ and in these units $k_F^{-1}\simeq32$. The pair correlation function of an ideal Fermi gas is also shown for comparison.}
\label{fig3:gr}
\end{figure}

The onset of instability can also be observed in the two-body radial distribution function of the gas. The functions  $g_2^{\alpha \beta}(r)$, with $\alpha$, $\beta=h$, $l$ are proportional to the probability of finding two particles of $\alpha$ and $\beta$ type at distance $r$. Their behavior at unitarity in the case $\kappa=1$ has been analyzed in Refs.~\cite{astra, lobo}. For like particles $g_2^{\alpha \alpha}(r)$ is very similar to that of an ideal Fermi gas, because Pauli exclusion principle largely overcomes the induced correlations mediated by interactions with the other component. Instead a large peak is formed in $g_2^{\alpha \beta}(r)$ at short distances for unlike particles due to the their strong attractive interactions. A similar behavior is expected in $g_2^{\text{hl}}(r)$ when $\kappa>1$. The tendency of cluster formation is instead visible in the distribution function of like particles. In Fig.~\ref{fig3:gr}, we report results for $g_2^{\text{hh}}(r)$ and $g_2^{\text{ll}}(r)$ obtained for mass ratios $\kappa=10$ and 14. The distribution function of light particles is insensitive to the value of $\kappa$ remaining close to the ideal gas result. On the contrary, $g_2^{\text{hh}}(r)$ shows a large peak at short distances for $\kappa=14$, indicating the formation of bound states involving one light particle. The Pauli exclusion principle brings eventually $g_2^{\text{hh}}$ down to zero on length scales smaller than the interaction range $R$.

\begin{figure}[t]
\centerline{
\includegraphics[width=0.6\linewidth,angle=-90]{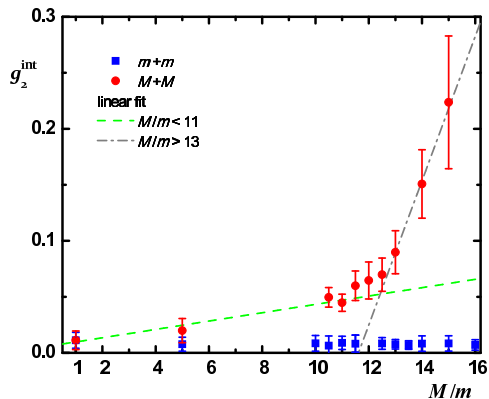}}
\caption{(color online): Results for the integrated distribution function $g_2^{\text{int}}(R_c)$ defined in Eq.~(\ref{g2int}) as a function of the mass ratio $\kappa$ for the case of \textit{ll} and \textit{hh} particles. In both cases, $R_c=8 R$.}
\label{fig4:int}
\end{figure}

An estimate of the critical mass ratio $\kappa_c$ characterizing the onset
of the cluster instability can be obtained by integrating the radial
distribution function up to a cut off length $R_c$,
\begin{equation}
g_2^{\text{int}}(R_c) = \frac{3}{R_c^3} \, \int_0^{R_c} g_2^{\alpha \alpha}(r) r^2 \ dr \ ,
\label{g2int}
\end{equation}
for the \textit{hh} and \textit{ll} distribution functions. The value of $R_c$ is chosen to be large compared to the range $R$ of the interatomic potential, giving the typical size of cluster states, but small compared to the mean interparticle distance.  In Fig.~\ref{fig4:int} we show the integrated values of $g_2^{\text{hh}}(r)$ and $g_2^{\text{ll}}(r)$, with $R_c=8 R$, as a function of the mass ratio. For the \textit{ll} distribution function the results of $g_2^{\text{int}}$ are essentially independent of $\kappa$, in agreement with the behavior shown in Fig.~\ref{fig3:gr}. On the contrary, in the \textit{hh} case, the corresponding $g_2^{\text{int}}$ increases linearly with the mass ratio up to $\kappa \simeq 11$. For $\kappa \gtrsim 13$ the increase is still linear, but the slope is significantly larger. In Fig.~\ref{fig4:int} we show the linear fit to the data in the regime $\kappa \leq 11$ and $\kappa \geq 13$. We checked that by changing the value of $R_c$, provided it remains within the range discussed above, the qualitative picture shown in Fig.~\ref{fig4:int} remains the same and in particular the value of $\kappa$ corresponding to the change of slope in $g_2^{\text{int}}(R_c)$ for the \textit{hh} particles.

\begin{figure}[t]
\includegraphics[width=0.9\linewidth,angle=0]{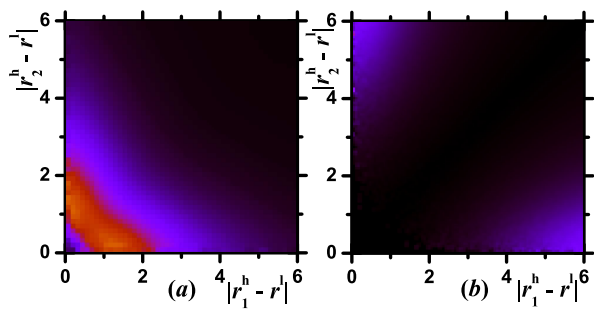}
\caption{Probability distribution as a function of the distances $|\bm{r}_1^{\text h} - \bm{r}^{\text l}|$ and $|\bm{r}_2^{\text h} -\bm{r}^{\text l}|$ for a three-particle system composed by two heavy and one light particles with mass ratio $\kappa=20$. \textit{(a)}: normal wave function. \textit{(b)}: superfluid wave function.}
\label{fig5:dens}
\end{figure}

An important question to understand is the apparent stability of simulations carried out using the superfluid guiding function (\ref{Pair}) when the mass ratio is larger than $\kappa_c$. In order to gain insight on that, we consider a simple system composed by two heavy and one light particles. We perform FN-DMC simulations of this three-body problem using as guiding functions $\Psi_T = \prod_{i=1,2}f(|{\bf r}_i^\textit{h}-{\bf r}^\textit{l}|)\sin(2\pi(x_1^\textit{h}-x_2^\textit{h})/L)$, corresponding to the JS normal wave function (\ref{JS}), and   $\Psi_T=\phi(|{\bf r}_1^\textit{h}-{\bf r}^\textit{l}|)-\phi(|{\bf r}_2^\textit{h}-{\bf r}^\textit{l}|)$, corresponding to the superfluid wave function (\ref{Pair}). The nodal surfaces bare important differences: the molecular orbital of the superfluid wave function is spherically symmetric while the one of the JS function possesses the cubic symmetry of plane waves in a box. Formation of a trimer bound state is strongly suppressed in the  case of the superfluid wave function. In fact, the two heavy particles repel each other due to the Pauli principle and prefer to stay apart, while the light particle attracts both heavy particles. If one assumes that in the trimer the heavy particles will preferentially stay at the same distance from the light particle, then a node appears in $\Psi_T$ when $|\bm{r}_1^{\text h} - \bm{r}^{\text l}| =|\bm{r}_2^{\text h} - \bm{r}^{\text l}|$. On the contrary, in the JS guiding function $\Psi_T$ there is an enhanced probability of arranging two heavy particles at the same distance from the light particle. This different behavior of the two wave functions is clearly shown in Fig.~\ref{fig5:dens} where we show the spatial dependence of the probability of finding the two heavy particles  at the distances $|\bm{r}_1^{\text h} - \bm{r}^{\text l}|$ and $|\bm{r}_2^{\text h} - \bm{r}^{\text l}|$ from the light particle. This correlation function vanishes when both distances approach zero according to the Pauli exclusion principle acting on the heavy particles. At the mass ratio $\kappa=20$ shown in the figure, one sees that the probability of finding the three particles close to each other is greatly increased in the case of the JS function. The peak corresponds to the sum $|\bm{r}_1^{\text h} -\bm{r}^{\text l}|+|\bm{r}_2^{\text h} -\bm{r}^{\text l}| \simeq R$. On the contrary, for the superfluid wave function the probability of finding $|\bm{r}_1^{\text h} -\bm{r}^{\text l}| = |\bm{r}_2^{\text h}-\bm{r}^{\text l}|$ is greatly suppressed and the overlap of this wave function with cluster states composed of two heavy and one light particle is therefore very small.

In conclusion, we investigate the stability at unitarity of a many-body mixture of two fermionic species with mass ratio $\kappa > 1$ using the FN-DMC method. Our results show that the mixture is superfluid and stable up to a critical mass ratio $\kappa_c=13\pm1$. For larger mass ratios, an instability sets in towards the formation of clusters. The study of the pair correlation function indicates that the relevant cluster states driving the instability are formed by one light particle and two or more heavy particles within distances on the order of the range $R$ of the potential. Within our approach, no many-body instability is observed for $\kappa<12$ where four and five-body bound states have been found in few-body calculations~\cite{blume2,carlson1}.

We acknowledge partial financial support from the DGI (Spain) Grant No.~FIS2011-25275 and Generalitat de Catalunya Grant No.~2009SGR-1003. G.E.A. acknowledges support from the Spanish MEC through the Ramon y Cajal fellowship program. S.G. acknowledges support by ERC through the QGBE grant.

\end{document}